# A simplified reconstruction of Positron Emission Tomography image using Time of Flight simulated data in Gate 10

Marcin Balcerzyk[1,2], Santiago Jimenez-Serrano[3], Óscar Pietrzyk[3], Edwing Ulin-Briseño[3,] Marta Freire[3].

1. Bioaraba Health Research Institute, New Technologies and Information Systems in Health Research Group, 01009 Vitoria-Gasteiz, Spain.
2. IKERBASQUE, Basque Foundation of Science, Plaza Euskadi 5, 48009 Bilbao, Spain.
3. Instituto de Instrumentación para Imagen Molecular, Centro mixto CSIC-Universitat Politècnica de València, 46022 Valencia, Spain.

| Author | ORCID |
| --- | --- |
| Marcin Balcerzyk[1,2] | 0000-0001-6030-7416 |
| Santiago Jiménez-Serrano[3] | 0000-0003-2917-6053 |
| Óscar Pietrzyk[3] | 0009-0006-2596-4940 |
| Edwig Ulin-Briseño[3] | 0009-0004-3921-8071 |
| Marta Freire[3] | 0000-0002-1150-383X |

## Novelty and significance

This manuscript introduces a simplified, rapid PET image TOF-driven reconstruction method requiring minimal software adaptation for new scanner geometries. Unlike existing packages, it enables direct, high-resolution 3D mapping from simulated coincidences—including precise TOF information—without artifacts and with sub-second processing. The approach is particularly significant for fast prototyping of evolving PET designs, as it allows easy analysis of image statistics, flexible spatial resolution and sensitivity calibration, and streamlining early-stage PET development. It should be noted that current PET detectors do not achieve the necessary time resolution of 3 ps or higher to fully leverage this method in real PET scanners.

# Abstract (192 words)

**Background:** Ultrafast Time-of-Flight (TOF) information in Gate 10 simulations enables direct 3D PET reconstruction without scanner-specific modeling. Although such picosecond timing is not achievable in current detectors, simulated data allow exploration of idealized TOF regimes and rapid evaluation of prototype PET designs.

**Methods:** The TOF-driven reconstruction assigns each coincidence to its annihilation position using sub-picosecond timestamps, producing voxelized 3D histograms with flexible voxel size and field-of-view selection. The approach operates directly on Gate-sorted coincidences, requires no corrections, and outputs MHD/DICOM images. Timestamp precision (~10^-13 s) enables localization on the millimeter scale.

**Results:** Full 3D images of the simulated INSPIRE PET scanner are generated in under one second. The method resolves 0.25 mm features in point-source studies and 1.2 mm rods in the Derenzo phantom, reproduces ground-truth activity distributions in image-quality tests, and maintains quantitative stability across geometries. Performance reflects the theoretical benefits of ultrafast TOF rather than current detector capabilities.

**Significance:** This fast, geometry-agnostic reconstruction tool supports early-stage PET prototyping, allowing rapid assessment of spatial resolution, sensitivity, and uniformity without implementing complex reconstruction software. It is broadly applicable to any simulated PET system with sorted coincidences and enables systematic exploration of ultrafast TOF performance in a controlled environment.

# Introduction

Positron Emission Tomography (PET) is an imaging technique that uses radiotracer labeled with positron-emitting isotopes, injected into patients, to map drug activity in the body over time. When a positron emitted from isotopes like F-18 or C-11 meets an electron in tissue, they annihilate, producing two gamma rays in opposite directions with a characteristic 511 keV energy each. Detectors around the subject capture these gamma rays, and if both are detected within a specific time frame, their points of detection define a line of response (LOR) where the positron-electron annihilation occurred. Images are reconstructed from a large number of these LORs.

If the detectors account for Time-of-Flight capabilities in the picosecond range, it becomes possible to delimit the region along the LOR where the annihilation took place. Current clinical PET scanners can reach a coincidence time resolution of approximately 200 ps, which restricts the localization of the annihilation event to about 120 mm [1].

If detectors achieve a coincidence time resolution of 10 ps or less, direct 3D reconstruction becomes feasible, since the TOF limits the event location to approximately 1.5 mm—similar to small animal PET scanners [2,3]. Some research detectors can reach about 100 ps, which is above the necessary threshold with the record of 32 ps and 4 mm spatial resolution presented by Kwon [4].

Before building a real PET scanner, its design and performance are typically simulated. Gate 10 [5-7] is a Python 3 library used for simulating gamma ray interactions and coincidence sorting, though it currently (as of version 10.0.3) lacks integrated reconstruction tools. Other packages like STIR [8], Castor [9], and PyTomograpy [10] exist, but they are generally standalone C++ programs or require detailed scanner geometry and specific data coding. This paper describes a streamlined Python-based approach for PET image reconstruction that reads coincidence events from a root file generated by Gate simulations and uses TOF information. The proposed method, named TOF-driven reconstruction method, is primarily intended for simulated data, as the required TOF resolution is not currently achievable with existing PET detector technology. The method utilizes both the spatial coordinates and highly precise timing information of each coincidence event to assign counts to individual voxels in the reconstructed image. TOF information is taken directly from the Gate simulation timestamps, which are available with subpicosecond numerical precision and thus defines coincidence time resolution. This allows us to investigate the reconstruction behavior in an idealized TOF regime (few ps range) unattainable with current systems. Our aim is not to emulate the performance of present detectors, but to quantify the potential of ultrafast TOF in a controlled and reproducible simulation environment. By employing a 64-bit floating-point variable for timestamping, the system achieves approximately 17 significant decimal digits of precision. This precision translates directly to a coincidence resolving time of $10^{-13}$ seconds.

The proposed method has been evaluated using simulated data from a PET prototype (INSPIRE) currently under construction. The paper presents reconstruction results for various phantom types—including point sources, a Derenzo-like phantom and a dedicated Image Quality (IQ) phantom—and compares these with the ground truth spatial distribution of isotope decays within the phantom.

# Materials and Methods

## INSPIRE PET scanner

The INSPIRE PET scanner has been simulated using Gate 10.0.3 on Linux[1]. The system geometry is made out of 16 semi-monolithic blocks based on BGO crystals (see Figure 1 A). Each block is constructed from 24 slabs of approximately 2 mm × 30 mm × 50 mm separated by enhanced specular reflectors (ESR). The detector blocks are mounted on a ring with a 260 mm inner diameter. The slabs are aligned parallel to the longitudinal axis of the scanner, generating a 50 mm axial field of view, see Figure 1 A. The data acquisition system features 17 ns coincidence window due to the long decay time of BGO luminescence. The energy resolution at 511 keV was set to 20%. The energy window is from 410 keV to 610 keV (±20%). Coincidences are identified in % of random and scattered in the sorting stage and the policy `removeMultiples` is used.

The INSPIRE scanner is used exclusively as a simulated detector geometry. The TOF-driven reconstruction presented in this work relies solely on the high-precision timestamps provided by GATE simulations and does not assume a detector with true picosecond-level timing. No experimental data from the physical INSPIRE system are included in this study.

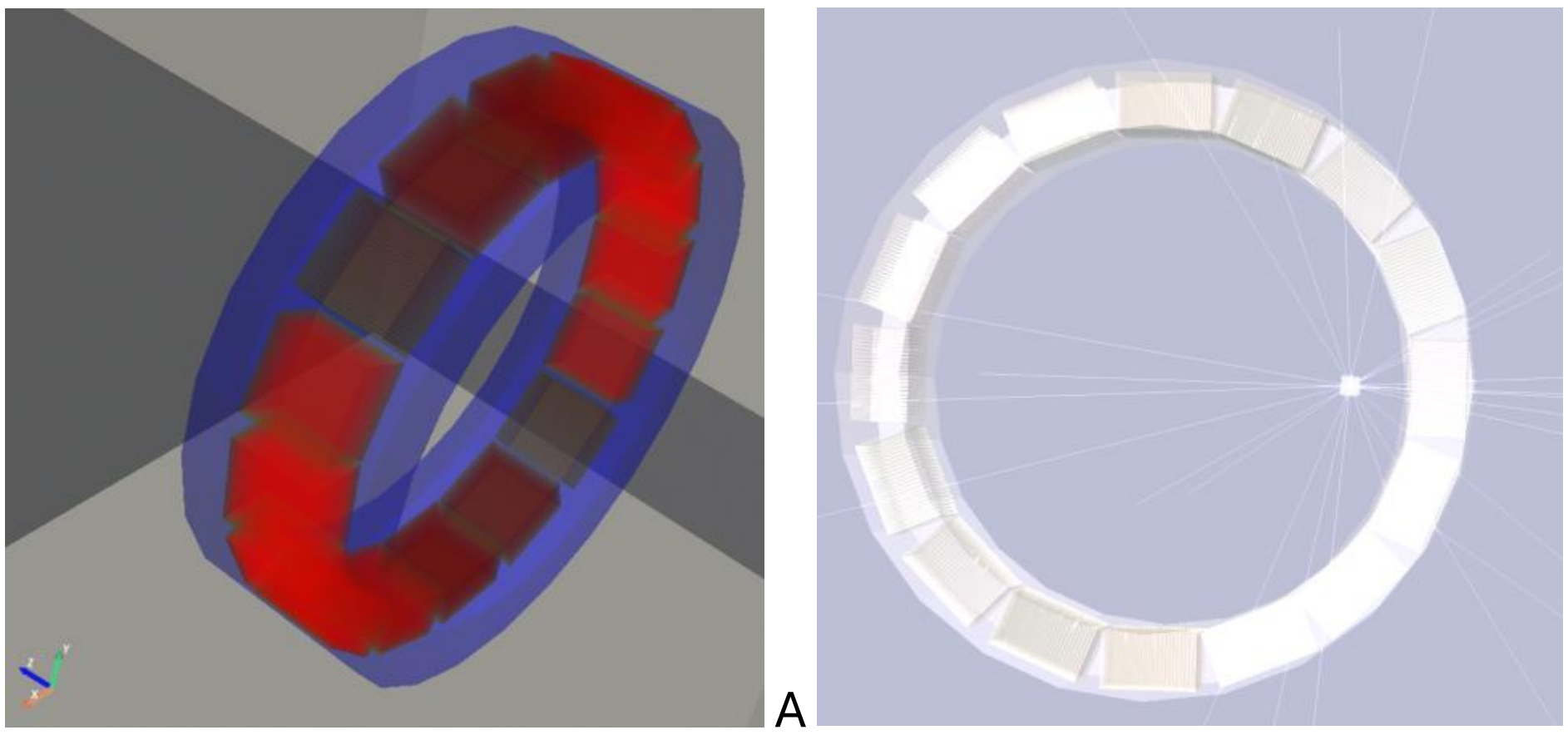

A B

*Figure 1 A.* ***Simulated INSPIRE PET scanner general view.*** *B. Simulated Na-22 point source in extreme radial position at x = 100 mm with tracks of 511 keV gamma-rays.*

## Simulated Phantoms

We followed the NEMA NU 4-2008 [11] protocol to evaluate the spatial resolution of the simulated scanner and TOF-driven reconstruction method.  We simulated $10^5$ counts from a 0.25 mm spherical source placed in a 10 mm PMMA cube, using an ion source of Na-22. The physics list was QGSP_BERT_HP. This list was necessary to include the decay of Na-22, diffusion and range of positrons resulting from disintegration of the nucleus. The source was moved radially across the Field Of View (FOV) at 5, 10, 15, 25, 50, 75, and 100 mm, and measurements were repeated at 12.5 mm axial distance from the center. The source was not placed at 125 mm off-radial position as it was already overlapping with the detector. Figure 1 B shows the simulated system with the source at x = 100 mm and z = 12.5 mm.

[1] Ubuntu 24.04.2 via the Windows 11 ARM64 WSL2 subsystem. Python 3.12 was employed on Linux, with development conducted in Visual Studio Code 1.103.0 utilizing Pylance v2025.7.1 and GitHub Copilot version 1.350, powered by GPT-4.1. GitHub Copilot facilitated the construction of the Python code.

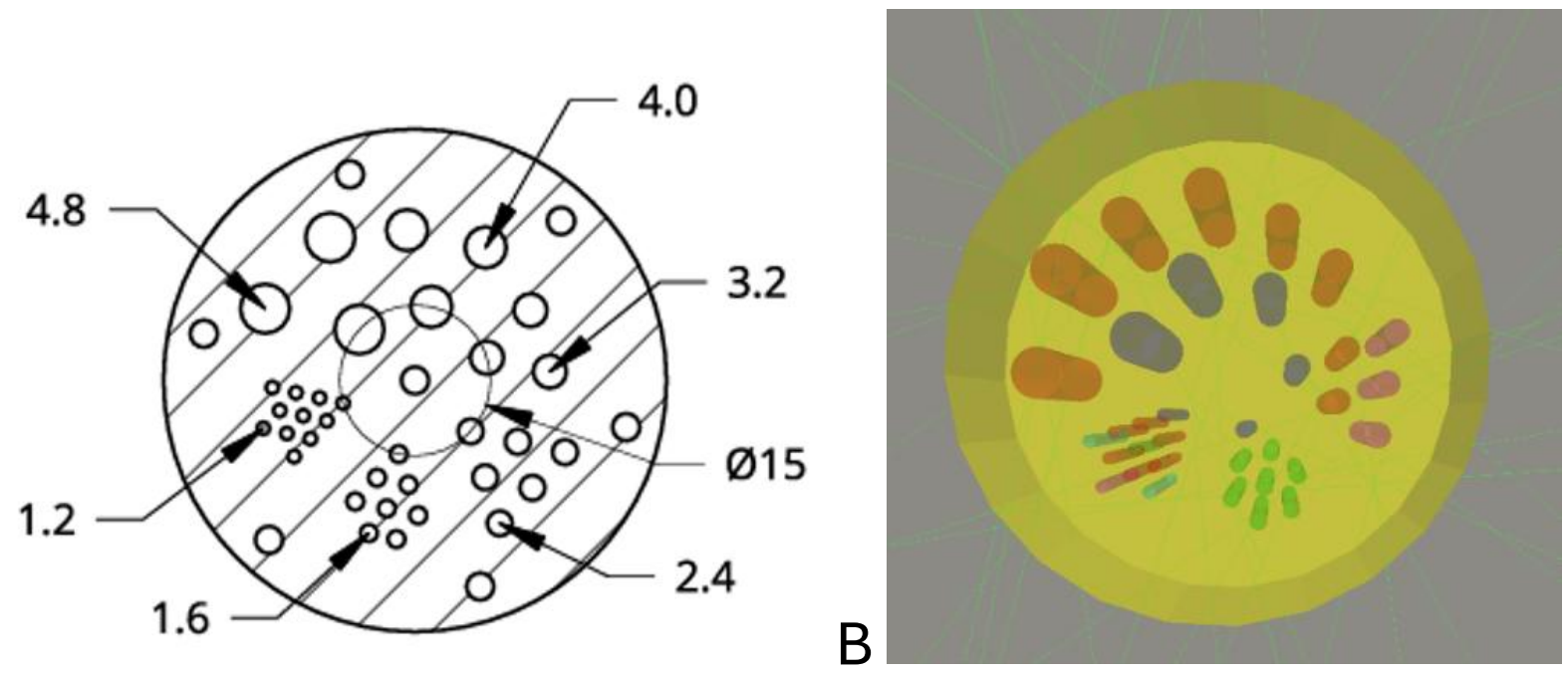

*Figure 2 A.* ***Sketch of the Derenzo-like phantom.*** *The phantom contains the holes for the liquid source, arranged in groups with diameters of 4.8, 4.0, 3.2, 2.4, 1.6, and 1.2 mm, with center-to-center separations equal to twice the diameter. The cylinder has a diameter of 50 mm. B. Derenzo-like phantom visualized in PyVista.*

A second simulated case was a custom Derenzo-like phantom, as provided by Phantech LLC (model D501248, Madison, WI, USA), with rods of 4.8, 4.0, 3.2, 2.4, 1.6 mm separated by twice the diameter, and 15 mm length (see Figure *2* A). PyVista rendering of the phantom is shown in Figure *2* B[2]. The rods were filled with a positron source with F-18 energy spectrum in water. Gate uses "e+" type source with selectable energy spectrum, which here is chosen as F-18 spectrum. Positrons are generated with appropriate energy and hence the range is simulated in the material they pass through, here water and PMMA. The physics list was G4EmStandardPhysics_option3. The Python program with a function for simulating the Derenzo-like phantom filled with 10 Bq/mL of F-18 is available as supplementary material RodPhantomAndSource[3]. 3.8 million coincidences were recorded.

The last simulated phantom is a dedicated Image Quality (IQ) phantom described in detail in Ref. [12]. The phantom of 139 mm diameter and 119 mm height had five cylindric hot regions of diameter 20, 15, 12, 9, 6 and 4.5 mm reaching half height of the phantom and the rest cold region filled with [F-18]FDG with the hot:cold concentration ratio of 4:1. The cold activity concentration was 10 kBq/mL.

---

[2] The original phantom is filled with a syringe to ensure a proper filling of all tubes from the largest diameter to the smallest with connecting tubing (this was not simulated).

[3] For the program to function properly, place additional supplemental files Phantech_Derenzo.py, user_colors.py in the same directory and GateMaterials.db into /data subdirectory. Visu.wrl is a VRML file with 3D rendering of the phantom and several gamma tracks.

## Phantom reconstruction and validation

We have implemented the TOF-driven reconstruction method adapting the mathematical framework introduced by Kwon *et al.* [4] was adapted to estimate the annihilation position in three dimensions using ps-level TOF information. For clarity, the equations presented here are only for the x-coordinate, while the corresponding expressions for the y- and z- coordinates are described in Appendix A.

Let 0 indicate the annihilation, 1 indicate first detected coincidence and 2 the second one, *t* time of that coincidence and *x* its position, $c_x$ is the projection of the speed of light on the *x* axis. The first equation in the system of equations indicates that the difference in position of both coincidences (right side) is equal to the speed of light component on the axis multiplied by the time between the coincidence and the time of annihilation. The second equation in the system of equations indicates that the position of annihilation on axis can be calculated from the position of the first coincidence plus the speed of light component on that axis multiplied by the difference between the time of the first coincidence minus the time of the annihilation.

**System of equations:**

$$c_x(t_1 - t_0) + c_x(t_2 - t_0) = x_2 - x_1$$

$$x_0 = x_1 + c_x(t_1 - t_0)$$

**Solution**

$$x_0 = x_1 + \frac{c_x(t_1 - t_2)}{2} + \frac{x_2 - x_1}{2}$$

*Equation 1*

The formulas used in Python are described in Appendix A.

Using a 64-bit floating-point variable for timestamps provides about 17 digits of precision, resulting in a coincidence resolving time of $10^{-13}$ seconds. To explain further, even for a simulated scan lasting up to 3 hours (about 10,800 seconds), the full range of timestamps can be represented in increments as small as $10^{-13}$ seconds, ensuring that every coincidence event within this duration is uniquely encoded. This prevents any loss of timing information due to rounding or overflow and safeguards the integrity of the coincidence resolving time throughout the scan. As a result, the voxel size can be freely adjusted according to the available coincidence statistics and the desired spatial sampling. The 3D region of interest for reconstructed image may be freely positioned anywhere within the scanner's field of view.

The simulated data from the point sources and phantoms were reconstructed using the TOF-driven reconstruction method. Typical execution of reconstruction and DICOM file takes 1-2 s depending on the coincidence file size, but these times were obtained for up to 300 MB coincidence root files and 2M coincidences. Notice that all the reconstructed images represent idealized performance under simulated picosecond timing and should be interpreted as illustrating the theoretical resolving power of TOF, rather than reflecting the behavior of existing PET scanners.

The simulated data from the Na-22 point sources were reconstructed using a voxel size of 0.25 mm. For the spatial resolution study, The Full Width at Half Maximum (FWHM) and Full Width at Tenth Maximum (FWTM) of the profiles in the radial, transaxial and axial directions were determined. The Derenzo-like phantom was reconstructed with a voxel size of 0.25 mm and 1 mm. The IQ phantom was reconstructed with a voxel size of 2 mm. The background variability (BV) and percent contrast (PC) values were obtained based on NEMA NU 4-2008 standard definitions [11] using the open-source software ChameleonIQ [13].

For each case, the reconstructed images and profiles were compared with the simulated location of the volume of the radioactivity source from the root file of the simulation. This way the simulated events in the phantom were recorded with precise location of each of them (i.e. not as a pixelated image) and could serve for comparison.

# Results

## Na-22 point source

Results of the spatial resolution simulation study are presented in Figure 3. The reconstructed image for the Na-22 point source located at x=100 mm and y=12.5 mm is shown in Figure 3 A, with no observable smudge or spread in any direction. The profiles for the transaxial direction (Y) of the spot source located at the center and at x=100 mm and y=12.5 mm are provided in Figure 3 B and C, respectively. The measured spatial resolution was consistent in all three directions and at all simulated positions, with a value of 0.25 mm. Similarly, the Full Width at Tenth Maximum (FWTM) was uniform resulting in 0.75 mm. There was no detected deterioration in spatial resolution, even though some gamma rays were not stopped in the empty areas between detectors, as illustrated in Figure 1 B. Figure 1 D and E show the ground truth profiles for the two source positions.

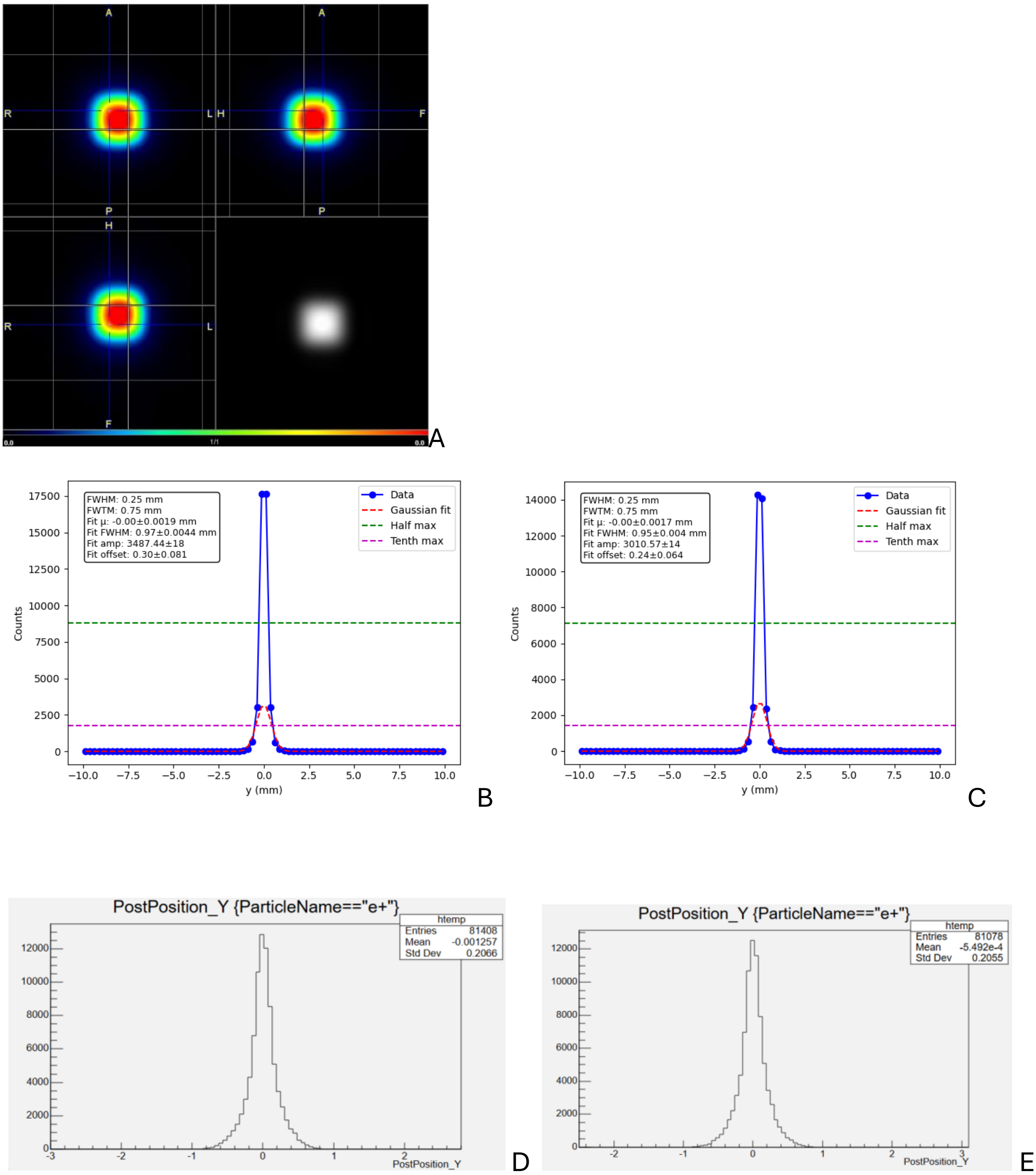

*Figure* 3 ***Spatial resolution study based on NEMA-2008.*** *A. Reconstructed image of the Na-22 point source located at x = 100 mm and z = 12.5 mm using the proposed TOF-driven reconstruction method. The grid measures 1 mm, with a voxel size of 0.25 mm. Maximum of cold color scale is at 12,058 coincidences, not zero as in the image. B. Profile of the Na-22 point source located at x = 0 mm, and z = 0 mm using the proposed reconstructor. The measured resolution is 0.25 mm in the transaxial direction. It remains consistent across all directions and positions. Voxel size is 0.25 mm. C. Profile of the Na-22 point source located at x = 100 mm and z = 12.5 mm using the proposed reconstructor. A spatial resolution of 0.25 mm was measured in the transaxial direction . D. Ground truth histogram of positron annihilation for Na-22 located at x = 0 mm and z = 0 mm . E. Ground truth histogram of positron annihilation for Na-22 located at x = 100 mm and z = 12.5 mm.*

## Figure 3Figure 3Figure 3Figure 1Derenzo-like phantom

Figure 4 displays reconstructions of the Derenzo-like phantom. In Figure 4 A (1 mm voxels), all holes down to 1.2 mm are visible, though activity diminishes near the longitudinal edges due to lower scanner efficiency. Figure 4 B (0.25 mm voxels) shows less difference in activity concentration between small poles, indicating reduced partial volume effects. Figure 4 C shows a profile through three rods of 1.2 mm diameter. Minor rises found between the main peaks are caused by positrons, which are originated from nearby poles and are scattered in PMMA. The simulated location of the volume of the F-18 source in the phantom (ground truth), including events in the PMMA region are shown in Figure 4  D. Figure 4 E shows a ground truth profile corresponding to Figure 4 C.

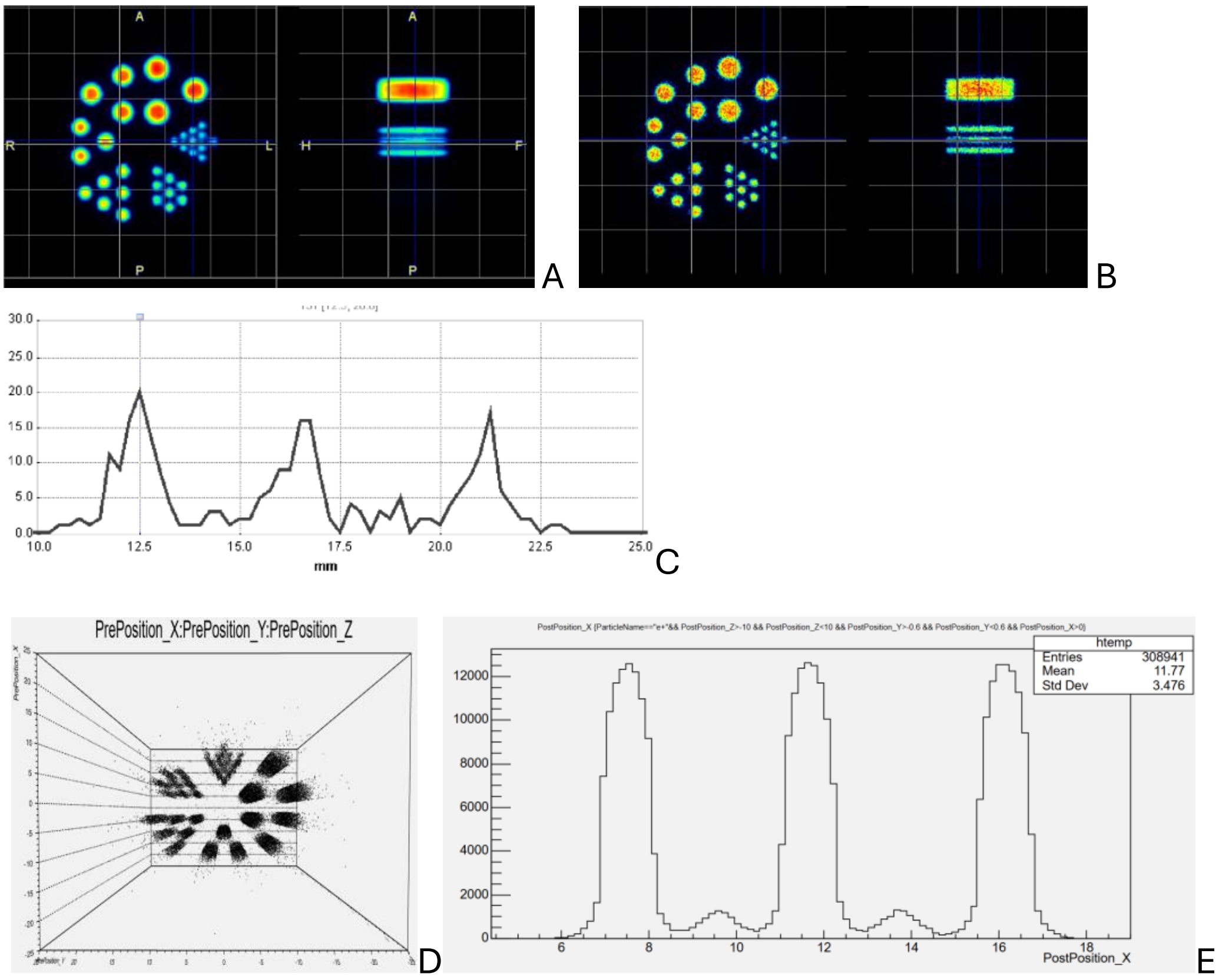

*Figure 4 A. Reconstruction of the Derenzo-like phantom using the proposed TOF-driven reconstruction method and 1 mm voxel size. Maximum of cold color scale is at 1501 coincidences, not zero as in the image. B. Reconstruction with 0.25 mm voxels from the same dataset, showing less partial volume effect.  All hole diameters—4.8 mm, 4.0 mm, 3.2 mm, 2.4 mm, 1.6 mm, and 1.2 mm—are clearly visible. Maximum of cold color scale is at 41 coincidences, not zero as in the image. C. Profile through 1.2 mm rods diameter. D. Simulated location of the volume of the F-18 source in the phantom, including events in the PMMA region. E. Ground truth profile corresponding to C of 1.2 mm rods.*

Figure 4 shows the reconstructed image of the IQ phantom together with the ROI configuration used to compute PC and BV for each rod size.

The PC and BV values are shown in Table 1. PC decreased progressively with insert size, from 55.6 ± 2.0% for the 20-mm insert down to 27.9 ± 9.6% for the smallest 4.5-mm insert, reflecting the expected loss of contrast recovery for smaller objects due to partial volume effects. Conversely, BV increased slightly as insert/background ROI size decreased, ranging from 7.2 ± 0.7% for the 20-mm ROI to 8.6 ± 0.8% for the 4.5-mm ROI, consistent with greater statistical noise in smaller sampling regions.  Table 1

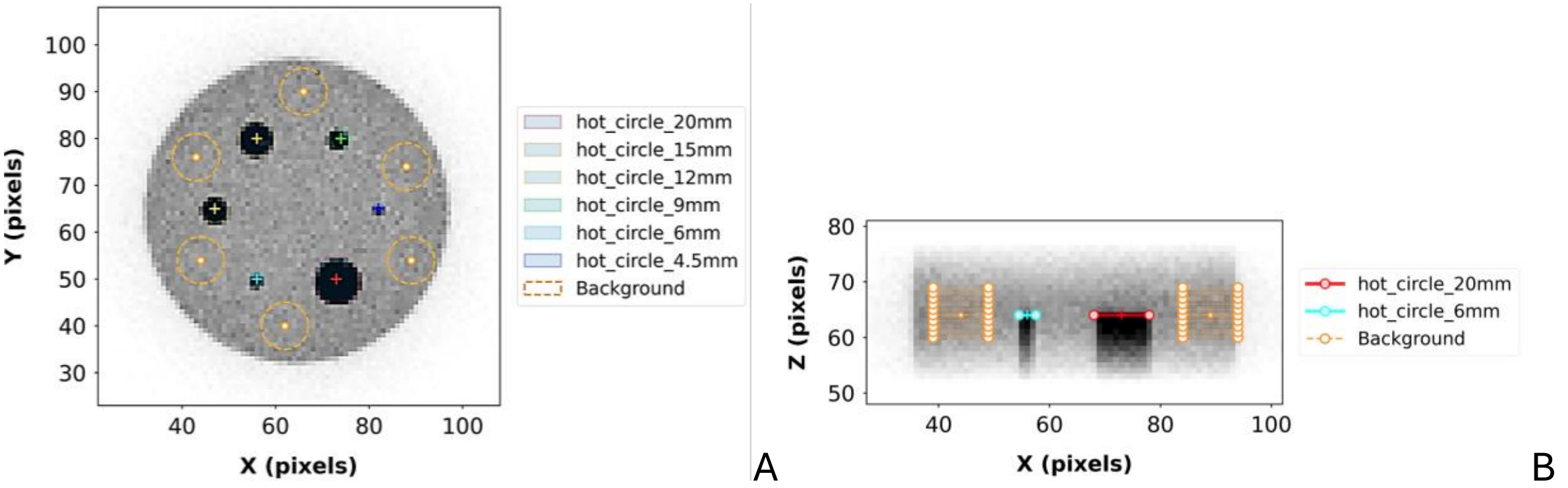

*Figure 5* ***Reconstructed Image Quality phantom with the regions of interest (ROI) used for percent contrast (PC) and background variability (BV) calculations.*** *(A) Transaxial slice of the six hot rods (diameters 20, 15, 12, 9, 6, and 4.5 mm), with colored crosses marking rods centers and dashed oranges circle indicating the 6 background ROIs. (B) Coronal view along the inserts axis, showing the 20 mm (red) and 6 mm (cyan) inserts with the background ROIs (orange), used to verify correct axial placement and sizing of the ROIs prior to measure.*

*Table 1 PC and BV values obtained for each hot rod of the Image Quality phantom, expressed as mean ± error*

| Spot size (mm) | 20 | 15 | 12 | 9 | 6 | 4.5 |
|---|---|---|---|---|---|---|
| Percent contrast (%) | 55.6±2.0 | 53.7±2.4 | 48.2±3.5 | 40.6±5.0 | 36.6±6.4 | 27.9±9.6 |
| Background variability (%) | 7.2±0.7 | 7.1±0.7 | 7.0±0.7 | 7.3±0.7 | 7.8±0.7 | 8.6±0.8 |

# Discussion and conclusions

This paper introduces a simple, fast and efficient method for PET image reconstruction of Gate simulated data using TOF information. While the approach is primarily applicable to simulated datasets due to the specific TOF requirements, it is particularly valuable during the initial design phases of PET equipment. Additionally, the method enhances accessibility to reconstruction algorithms, making them usable for professionals without programming expertise. STIR, PyTomography and CASToR are full reconstruction packages that use all corrections and require substantial effort to implement an existing PET scanner design. Our method requires only a root file from the GATE simulation with sorted coincidences and no data from the scanner geometry.

The TOF-driven reconstruction method described here is especially useful for rapid PET scanner prototyping, providing qualitative information that supports design improvements. It can be applied to any PET scanner system with sorted coincidence data. Reconstruction and DICOM file saving typically require just 1–2 seconds, depending on the root file size. Each voxel delivers detailed data on the number of detected coincidences, allowing flexible adjustments to decay counts, simulation duration, or voxel size as needed. High time resolution in this method demonstrates the potential for achieving ultra-high, 1 mm resolution in clinical PET systems, contingent upon the development of suitable 511 keV gamma detectors.

A potential point of confusion is the coexistence of simulated TOF information with detector geometries whose real counterparts lack TOF capability. For clarity, the TOF used in this work is entirely numerical, provided by Gate at picosecond precision, and does not correspond to the physical timing resolution of current detectors. Our objective is to evaluate how an idealized TOF regime could support rapid, geometry-agnostic histogram-based image reconstruction. Therefore, we intentionally do not compare against experimental TOF data, as no existing system provides the required timing resolution. Instead, the method benchmarks the intrinsic limits of TOF-enabled 3D localization, providing a reference point for future detector development.

While TOF PET provides improved localization of the annihilation event by measuring the difference in arrival times of the coincident photons, its practical resolution is currently limited to approximately 0.5–5 ns in typical clinical systems. This constraint means that the reconstructed position of the annihilation cannot be pinpointed with sub-millimeter accuracy, and the contribution from scattered events is only reduced, not eliminated. As a result, TOF PET scanners are unable to fully resolve the spatial distribution of activity, particularly in regions with high scatter or complex geometries, highlighting a fundamental limitation in current TOF-based reconstruction approaches.

In the present manuscript, TOF is used directly to calculate the position of annihilation, demonstrating the potential improvement in spatial resolution. However, in conventional PET systems, the TOF limit restricts how accurately the annihilation site can be localized, and thus, scatter correction remains incomplete. This underscores the need for further advances in detector time resolution to achieve truly high-precision reconstruction and minimize scatter-related artifacts.

Compared to the 4 mm resolution at 32 ps reported by [4], histogram reconstruction with sub-picosecond time resolution and 0.25 mm spatial resolution used in Equation 1 approach shows promise for clinical whole-body PET scanners but is not yet suitable for breast, brain, or small animal scanners. 32 ps time resolution resulting in 4 mm spatial resolution is not sufficient for smaller diameter scanners where spatial resolution using conventional method is already below 1 mm. Higher time resolution of PET detectors is still required for practical applications, though this paper demonstrates the current limits of TOF reconstruction.

## Acknowledgements

The project was funded in part by Generalitat Valenciana (Spain), project number INNVA1/2024/72. Edwing Ulin-Briseño was supported by the Santiago Grisolias of the Generalitat Valenciana Grant No. CIGRIS/2023/029. Marta Freire was supported by the CIAPOS Program for Researchers in Postdoctoral Phase of the Generalitat Valenciana under Grant No. CIAPOS/2023/141.

# Appendix A

In the simulated root file of coincidences the $x_1$ is denoted by PostPosition_X1 and $t_1$ by GlobalTime1. The time units are ns, so the speed of light is coded as 299.792 mm/ns in the following formulas. The length units are mm.

Formulas for reconstruction of X, Y and Z coordinates of the annihilation event derived after simplifying Equation 1:

```
X      (PostPosition_X2+PostPosition_X1)/2.0+299.792*(GlobalTime1-
GlobalTime2)*(PostPosition_X2-PostPosition_X1)/2.0/sqrt((PostPosition_X2-
PostPosition_X1)**2+(PostPosition_Y2-PostPosition_Y1)**2+(PostPosition_Z2-
PostPosition_Z1)**2)
```

```
Y       (PostPosition_Y2+PostPosition_Y1)/2.0+299.792*(GlobalTime1-
GlobalTime2)*(PostPosition_Y2-PostPosition_Y1)/2.0/sqrt((PostPosition_X2-
PostPosition_X1)**2+(PostPosition_Y2-PostPosition_Y1)**2+(PostPosition_Z2-
PostPosition_Z1)**2)
```

```
 Z     (PostPosition_Z2+PostPosition_Z1)/2.0+299.792*(GlobalTime1-
GlobalTime2)*(PostPosition_Z2-PostPosition_Z1)/2.0/sqrt((PostPosition_X2-
PostPosition_X1)**2+(PostPosition_Y2-PostPosition_Y1)**2+(PostPosition_Z2-
PostPosition_Z1)**2)
```

These equations are used in the Python function BuildHistogram4Image with the following parameters:

```
def BuildHistogram4Image(output_path, root_filename, coincidences_tree, spacing_x,
spacing_y, spacing_z, range_minX, range_maxX, range_minY, range_maxY, range_minZ,
range_maxZ):
```

*Output_path* is a path to the root file (*root_filename*) where the tree of *coincidences* are stored from coincidence_sorter Gate internal function. *spacing_x* is the size of the voxel in *x* direction in mm, similarly for *y,z*. *range_minX* is the lower value of the image border in the coordinates used to store coincidences, similarly for maximum in *range_maxX* and *y,z* axes. The function requires the following libraries: opengate, uproot, numpy, awkward, SimpleITK, matplotlib, pathlib, scipy. The Python program is available in supplementary data as *BuildImage.py*. Note, that this function does not require any prior definition of the PET scanner.

The $x_0$ position data calculated from Equation 1 between *range_minX* and *range_maxX* are included in the 3D histogram of the bin size (*range_maxX* - *range_minX)/spacing_x* and similarly for *y,z* axes. The function stores image in MHD, NIFTi and DICOM formats using SimpleITK in one 3D file and in sliced format. No smoothing is used. We have checked the import of the DICOM data with ITK-SNAP program and all three versions were imported correctly. PMOD 4.5 did not import correctly pydicom saved 3D image.

# Supplementary material

## Python function BuildHistogram4Image

```
def BuildHistogram4Image(output_path, root_filename, coincidences_tree,
                         spacing_x, spacing_y, spacing_z, range_minX,
range_maxX,range_minY,range_maxY, range_minZ, range_maxZ,
                         from_time, to_time):
    # the procedure assumes "concidences" tree in the root file,
    # numbins_x is the number of pixels in the x y and z direction, spacing_x
is the size of each pixel in mm,
    # range_x is the range of the image in mm : (range_minX, rangemaxX),
(range_minY, range_maxY), (range_minZ, range_maxZ)
    print(f"Opening {output_path} / {root_filename} ...")
    root_file = uproot.open(root_filename)

    # Access the coincidences tree
    tree = root_file[coincidences_tree]

    # Extract data from a specific branch (e.g., energy values)
    X2 = tree["PostPosition_X2"].array()
    X1 = tree["PostPosition_X1"].array()
    Y2 = tree["PostPosition_Y2"].array()
    Y1 = tree["PostPosition_Y1"].array()
    Z2 = tree["PostPosition_Z2"].array()
    Z1 = tree["PostPosition_Z1"].array()
    T2 = tree["GlobalTime2"].array()
    T1 = tree["GlobalTime1"].array()

    # Filter events within the time range
    mask = (T1 >= from_time) & (T1 <= to_time) & (T2 >= from_time) & (T2 <=
to_time)
    T2 = T2[mask]
    T1 = T1[mask]
    X2 = X2[mask]
    X1 = X1[mask]
    Y2 = Y2[mask]
```

```
    Y1 = Y1[mask]
    Z2 = Z2[mask]
    Z1 = Z1[mask]

    # Apply a formula (example: square root transformation)
    processed_dataX=(X2+X1)/2.0+299.792*(T2-T1)*(X2-X1)/2.0/((X2-X1)*(X2-
X1)+(Y2-Y1)*(Y2-Y1)+(Z2-Z1)*(Z2-Z1))**(1/2)
    processed_dataY=(Y2+Y1)/2.0+299.792*(T2-T1)*(Y2-Y1)/2.0/((X2-X1)*(X2-
X1)+(Y2-Y1)*(Y2-Y1)+(Z2-Z1)*(Z2-Z1))**(1/2)
    processed_dataZ=(Z2+Z1)/2.0+299.792*(T2-T1)*(Z2-Z1)/2.0/((X2-X1)*(X2-
X1)+(Y2-Y1)*(Y2-Y1)+(Z2-Z1)*(Z2-Z1))**(1/2)
    R=np.sqrt(processed_dataX**2+processed_dataY**2)
    # R is a radial histogram, not used in this code
    # Create a 3D histogram
    #num_bins_x, num_bins_y, num_bins_z = 260, 260, 50  # Adjust as needed
    #num_bins_x, num_bins_y, num_bins_z = 260*2, 260*2, 50*2  # Adjust as
needed HD
        #num_bins_x, num_bins_y, num_bins_z = 100, 100, 100  # Adjust as
needed HDLzoom

    x_positions = processed_dataX
    y_positions = processed_dataY
    z_positions = processed_dataZ
    # in the line below change (-10,10) to (-range_x, range_x) etc. to get the
histogram in the correct range
    num_bins_x, num_bins_y, num_bins_z=int((range_maxX-range_minX)/spacing_x),
int((range_maxY-range_minY)/spacing_y), int((range_maxZ-range_minZ)/spacing_z)

    hist_3d, edges = np.histogramdd((x_positions, y_positions, z_positions),
bins=(num_bins_x, num_bins_y, num_bins_z), range=((range_minX,range_maxX),
(range_minY,range_maxY), (range_minZ,range_maxZ)))
    hist_3d_np = np.array(ak.to_list(hist_3d))
    hist_3d_np = hist_3d_np.reshape((num_bins_x, num_bins_y, num_bins_z))

    # Transpose the NumPy array to match SimpleITK's (x, y, z) order
    hist_3d_np = np.transpose(hist_3d_np, (2, 1, 0))
    hist_3d_np = hist_3d_np.astype(np.uint16)  # Convert to uint16

    # Get the size of the image
    image_size_x, image_size_y, image_size_z = hist_3d_np.shape  # (x, y, z)
    # spacing_x, spacing_y, spacing_z = 0.1, 0.1, 0.1  # Spacing in mm

    # Calculate the center of the image
    center_x = - (image_size_x * spacing_x) / 2
    center_y = - (image_size_y * spacing_y) / 2
    center_z = - (image_size_z * spacing_z) / 2
```

```
    # Convert to SimpleITK image
    image = sitk.GetImageFromArray(hist_3d_np)

    # Set spacing, origin, and direction
    image.SetSpacing((spacing_x, spacing_y, spacing_z))  # Set the spacing in
mm
    image.SetOrigin((center_x, center_y, center_z))  # Set the origin to the
center
    image.SetDirection((1.0, 0.0, 0.0, 0.0, 1.0, 0.0, 0.0, 0.0, 1.0))  # Set
the direction cosines
    # image.

    # Save the image
    base_name =
f"{Path(root_filename).stem}_{from_time/60e9:.0f}min_{to_time/60e9:.0f}min"
    sitk.WriteImage(image, output_path / f"{base_name}.mhd")
    sitk.WriteImage(image, output_path / f"{base_name}ITK.dcm")
    sitk.WriteImage(image, output_path / f"{base_name}.nii")
```

## Python function RodPhantomAndSource

```
import matplotlib.pyplot as plt
import numpy as np
import opengate as gate
import pathlib
import pyvista
import SimpleITK as sitk
from user_colors import colors


current_path = pathlib.Path(__file__).parent.resolve()
data_path = current_path / "data"
output_path = current_path / "output7a"
# output_file = output_path / "dose.mhd"
# units
m = gate.g4_units.m
cm = gate.g4_units.cm
mm = gate.g4_units.mm
um = gate.g4_units.um
s = gate.g4_units.s
keV = gate.g4_units.keV
MeV = gate.g4_units.MeV
Bq = gate.g4_units.Bq # this is 1e-9 in utility.py

deg = gate.g4_units.deg
```

```
def MiniDeluxePhantom(sim, activity_concentration, visu=False,
run_timing_intervals=None):
    # Phantom cylinder of PMMA that contains the source
    # The function creates a MiniDeluxe phantom with a central PMMA cylinder
and multiple water Tubs around it, each containing a source of beta+
particles.
    # The sources are arranged in a hexagonal pattern around the phantom, with
additional sources at specific angles and distances.
    #   return phantom, water4sources, water4sourcesTubs
    # function returns phanton is the object with sources in water4sources and
poles of water in water4sourcesTubs.
    ui = sim.user_info
    phantom = sim.add_volume("Tubs", "phantom")
    phantom.rmax = 25 * mm
    # was 25 mm for Phantech
    phantom.rmin = 0 * mm
    phantom.dz = 50/2 * mm # was 35 for Phantech
    phantom.translation = [0 * mm, 0 * mm, 0 * mm]
    phantom.material = "PMMA"
    phantom.color = colors.yellow

    water4sourcesTubs=[]
    water4sources=[]
    # Tubs_radius_from_diam=[240/2] # for one cylinder
    Tubs_radius_from_diam=[1.2/2, 1.6/2, 2.4/2, 3.2/2, 4.0/2, 4.8/2]

    # Create six water Tubs with specified diameters and add corresponding
sources
    for i, diam in enumerate(Tubs_radius_from_diam):
        # Place each tub at a different angle around the phantom
        angle = i * (360 / len(Tubs_radius_from_diam))
        x = 1*15/2 * np.cos(np.deg2rad(angle)) * mm
        y = 1*15/2 * np.sin(np.deg2rad(angle)) * mm
        water_tub_rmax = diam * mm
        water_tub_dz = 50/2 * mm
        # For i != 0, place the tub at a distance of 15 mm from the center of
the phantom
        water_tub = sim.add_volume("Tubs", f"water4source_{i}")
        water_tub.mother = phantom.name
        water_tub.rmax = water_tub_rmax
        water_tub.rmin = 0 * mm
        water_tub.dz = water_tub_dz
        water_tub.translation = [x, y, 0 * mm]
        water_tub.material = "Water"
        water_tub.color = colors.blue
        # if i!= 0:


        if i == 0:
```

```
            # For i == 0, build a hexagon with a central pole (similar to i ==
1)
            # Change the name of the source for i==0 to avoid conflict with
i==1
            source_name = "beta+source_0_unique"
            hex_radius = 4 * water_tub.rmax
            hex_center_x = x + np.sqrt(3) * hex_radius *
np.cos(np.deg2rad(angle))
            hex_center_y = y + np.sqrt(3) * hex_radius *
np.sin(np.deg2rad(angle))
            # Add a blue pole in the center of the hexagon
            hex_center_tub = sim.add_volume("Tubs",
f"water4source_hex_center_{i}")
            hex_center_tub.mother = phantom.name
            hex_center_tub.rmax = water_tub.rmax
            hex_center_tub.rmin = 0 * mm
            hex_center_tub.dz = water_tub.dz
            hex_center_tub.translation = [hex_center_x, hex_center_y, 0 * mm]
            hex_center_tub.material = "Water"
            hex_center_tub.color = colors.green
            # Add the hex center tub to the list
            water4sourcesTubs.append(hex_center_tub)
            # Add a source for the center pole
            hex_center_source = sim.add_source("GenericSource",
f"beta+source_hex_center_{i}")
            hex_center_source.particle = "e+"
            hex_center_source.energy.type = "F18"
            hex_center_source.position.type = "cylinder"
            hex_center_source.position.radius = water_tub.rmax
            hex_center_source.position.dz = 15/2 * mm
            hex_center_source.position.translation = [hex_center_x,
hex_center_y, 0 * mm]
            hex_center_source.volume = np.pi *
(hex_center_source.position.radius ** 2) * 2 * hex_center_source.position.dz
            hex_center_source.energy.mono = 511 * gate.g4_units.keV
            hex_center_source.direction.type = "iso"
            hex_center_source.direction.theta = [0 * gate.g4_units.deg, 180 *
gate.g4_units.deg]
            hex_center_source.direction.phi = [0, 360 * gate.g4_units.deg]
            hex_center_source.activity = activity_concentration *
hex_center_source.volume / ui.number_of_threads * gate.g4_units.Bq
            water4sources.append(hex_center_source)
            hex_angle_offset = angle + 30

            # Add two tubs at the distal part of the hexagon, to the right and
left of the base
            # Move the left one two tubs diameters to the left, and the right
one two diameters to the right
```

```
            distal_angle = angle  # direction away from phantom center
            for k, offset in enumerate([-30, 30]):
                base_angle = distal_angle + offset
                # Calculate the base position
                base_x = hex_center_x + hex_radius *
np.cos(np.deg2rad(base_angle))
                base_y = hex_center_y + hex_radius *
np.sin(np.deg2rad(base_angle))
                # Perpendicular direction: distal_angle + 90 for right, -90
for left
                perp_angle = distal_angle + 90 if k == 1 else distal_angle -
90
                shift = 2 * 2 * water_tub.rmax  # 2 diameters
                base_x += shift * np.cos(np.deg2rad(perp_angle))
                base_y += shift * np.sin(np.deg2rad(perp_angle))
                base_tub = sim.add_volume("Tubs",
f"water4source_triangle_base_{i}_{k}")
                base_tub.mother = phantom.name
                base_tub.rmax = water_tub.rmax
                base_tub.rmin = 0 * mm
                base_tub.dz = water_tub.dz
                base_tub.translation = [base_x, base_y, 0 * mm]
                base_tub.material = "Water"
                base_tub.color = colors.cyan  # changed to cyan
                water4sourcesTubs.append(base_tub)
                # Add a source for each triangle base tub
                base_source = sim.add_source("GenericSource",
f"beta+source_triangle_base_{i}_{k}")
                base_source.particle = "e+"
                base_source.energy.type = "F18"
                base_source.position.type = "cylinder"
                base_source.position.radius = water_tub.rmax
                base_source.position.dz = 15/2 * mm
                base_source.position.translation = [base_x, base_y, 0 * mm]
                base_source.volume = np.pi * (base_source.position.radius **
2) * 2 * base_source.position.dz
                base_source.energy.mono = 511 * gate.g4_units.keV
                base_source.direction.type = "iso"
                base_source.direction.theta = [0 * gate.g4_units.deg, 180 *
gate.g4_units.deg]
                base_source.direction.phi = [0, 360 * gate.g4_units.deg]
                base_source.activity = activity_concentration *
base_source.volume / ui.number_of_threads * gate.g4_units.Bq
                water4sources.append(base_source)
            for h in range(6):
                hex_angle = hex_angle_offset + h * 60
                hex_x = hex_center_x + hex_radius *
np.cos(np.deg2rad(hex_angle))
```

```
                hex_y = hex_center_y + hex_radius *
np.sin(np.deg2rad(hex_angle))
                hex_tub = sim.add_volume("Tubs", f"water4source_hex_{i}_{h}")
                hex_tub.mother = phantom.name
                hex_tub.rmax = water_tub.rmax
                hex_tub.rmin = 0 * mm
                hex_tub.dz = water_tub.dz
                hex_tub.translation = [hex_x, hex_y, 0 * mm]
                hex_tub.material = "Water"
                hex_tub.color = colors.red
                water4sourcesTubs.append(hex_tub)
                # Add a source for each hex tub
                hex_source = sim.add_source("GenericSource",
f"beta+source_hex_{i}_{h}")
                hex_source.particle = "e+"
                hex_source.energy.type = "F18"
                hex_source.position.type = "cylinder"
                hex_source.position.radius = water_tub.rmax
                hex_source.position.dz = 15/2 * mm
                hex_source.position.translation = [hex_x, hex_y, 0 * mm]
                hex_source.volume = np.pi * (hex_source.position.radius ** 2)
* 2 * hex_source.position.dz
                hex_source.energy.mono = 511 * gate.g4_units.keV
                hex_source.direction.type = "iso"
                hex_source.direction.theta = [0 * gate.g4_units.deg, 180 *
gate.g4_units.deg]
                hex_source.direction.phi = [0, 360 * gate.g4_units.deg]
                hex_source.activity = activity_concentration *
hex_source.volume / ui.number_of_threads * gate.g4_units.Bq
                water4sources.append(hex_source)
            # Add the original first pole for i==0
            water4sourcesTubs.append(water_tub)
            source = sim.add_source("GenericSource", f"beta+source_{i}")
            source.particle = "e+"
            source.energy.type = "F18"
            source.energy.mono = 511 * gate.g4_units.keV
            source.position.type = "cylinder"
            source.position.radius = diam * mm
            source.position.dz = 15/2 * mm
            source.position.translation = [x, y, 0 * mm]
            source.volume = np.pi * (source.position.radius ** 2) * 2 *
source.position.dz
            source.energy.mono = 511 * gate.g4_units.keV
            source.direction.type = "iso"
            source.direction.theta = [0 * gate.g4_units.deg, 180 *
gate.g4_units.deg]
            source.direction.phi = [0, 360 * gate.g4_units.deg]
```

```
            source.activity = activity_concentration * source.volume /
ui.number_of_threads * gate.g4_units.Bq
            water4sources.append(source)

            # Place the extra tub further away in +x, maintaining two
diameters from the hexagon
            # Find the furthest hexagon tub in +x direction
            hex_xs = [hex_center_x + hex_radius *
np.cos(np.deg2rad(hex_angle_offset + h * 60)) for h in range(6)]
            max_hex_x = max(hex_xs)
            # Place new tub at max_hex_x + 2*2*water_tub.rmax (2 diameters
away from hexagon)
            new_x = max_hex_x + 2 * 2 * water_tub.rmax
            new_y = 0 * mm
            new_z = 0 * mm
            extra_tub = sim.add_volume("Tubs",
f"water4source_extra_highx_{i}")
            extra_tub.mother = phantom.name
            extra_tub.rmax = water_tub.rmax
            extra_tub.rmin = 0 * mm
            extra_tub.dz = water_tub.dz
            extra_tub.translation = [new_x, new_y, new_z]
            extra_tub.material = "Water"
            extra_tub.color = colors.magenta
            water4sourcesTubs.append(extra_tub)
            # Add a source for the extra tub
            extra_source = sim.add_source("GenericSource",
f"beta+source_extra_highx_{i}")
            extra_source.particle = "e+"
            extra_source.energy.type = "F18"
            extra_source.position.type = "cylinder"
            extra_source.position.radius = water_tub.rmax
            extra_source.position.dz = 15/2 * mm
            extra_source.position.translation = [new_x, new_y, new_z]
            extra_source.volume = np.pi * (extra_source.position.radius ** 2)
* 2 * extra_source.position.dz
            extra_source.energy.mono = 511 * gate.g4_units.keV
            extra_source.direction.type = "iso"
            extra_source.direction.theta = [0 * gate.g4_units.deg, 180 *
gate.g4_units.deg]
            extra_source.direction.phi = [0, 360 * gate.g4_units.deg]
            extra_source.activity = activity_concentration *
extra_source.volume / ui.number_of_threads * gate.g4_units.Bq
            water4sources.append(extra_source)
        if i == 1:
            # Build a hexagon at the same radius as the extra red poles (4 *
water_tub.rmax from center)
            hex_radius = 4 * water_tub.rmax
```

```
            # Move the whole green structure for i==1 by sqrt(3)*hex_radius
away from the center of the phantom
            hex_center_x = x + np.sqrt(3) * hex_radius *
np.cos(np.deg2rad(angle))
            hex_center_y = y + np.sqrt(3) * hex_radius *
np.sin(np.deg2rad(angle))
            # Add a green pole in the center of the green hexagon
            hex_center_tub = sim.add_volume("Tubs",
f"water4source_hex_center_{i}")
            hex_center_tub.mother = phantom.name
            hex_center_tub.rmax = water_tub.rmax
            hex_center_tub.rmin = 0 * mm
            hex_center_tub.dz = water_tub.dz
            hex_center_tub.translation = [hex_center_x, hex_center_y, 0 * mm]
            hex_center_tub.material = "Water"
            hex_center_tub.color = colors.green
            water4sourcesTubs.append(hex_center_tub)
            # Add a source for the center green pole
            hex_center_source = sim.add_source("GenericSource",
f"beta+source_hex_center_{i}")
            hex_center_source.particle = "e+"
            hex_center_source.energy.type = "F18"
            hex_center_source.position.type = "cylinder"
            hex_center_source.position.radius = water_tub.rmax
            hex_center_source.position.dz = 15/2 * mm
            hex_center_source.position.translation = [hex_center_x,
hex_center_y, 0 * mm]
            hex_center_source.volume = np.pi *
(hex_center_source.position.radius ** 2) * 2 * hex_center_source.position.dz
            hex_center_source.energy.mono = 511 * gate.g4_units.keV
            hex_center_source.direction.type = "iso"
            hex_center_source.direction.theta = [0 * gate.g4_units.deg, 180 *
gate.g4_units.deg]
            hex_center_source.direction.phi = [0, 360 * gate.g4_units.deg]
            hex_center_source.activity = activity_concentration *
hex_center_source.volume / ui.number_of_threads * gate.g4_units.Bq
            water4sources.append(hex_center_source)
            hex_angle_offset = angle + 30  # align with previous extra poles
            for h in range(6):
                hex_angle = hex_angle_offset + h * 60  # 6 vertices, 60 deg
apart
                hex_x = hex_center_x + hex_radius *
np.cos(np.deg2rad(hex_angle))
                hex_y = hex_center_y + hex_radius *
np.sin(np.deg2rad(hex_angle))
                hex_tub = sim.add_volume("Tubs", f"water4source_hex_{i}_{h}")
                hex_tub.mother = phantom.name
                hex_tub.rmax = water_tub.rmax
```

```
                hex_tub.rmin = 0 * mm
                hex_tub.dz = water_tub.dz
                hex_tub.translation = [hex_x, hex_y, 0 * mm]
                hex_tub.material = "Water"
                hex_tub.color = colors.green
                water4sourcesTubs.append(hex_tub)
                # Add a source for each hex tub
                hex_source = sim.add_source("GenericSource",
f"beta+source_hex_{i}_{h}")
                hex_source.particle = "e+"
                hex_source.energy.type = "F18"
                hex_source.position.type = "cylinder"
                hex_source.position.radius = water_tub.rmax
                hex_source.position.dz = 15/2 * mm
                hex_source.position.translation = [hex_x, hex_y, 0 * mm]
                hex_source.volume = np.pi * (hex_source.position.radius ** 2)
* 2 * hex_source.position.dz
                hex_source.energy.mono = 511 * gate.g4_units.keV
                hex_source.direction.type = "iso"
                hex_source.direction.theta = [0 * gate.g4_units.deg, 180 *
gate.g4_units.deg]
                hex_source.direction.phi = [0, 360 * gate.g4_units.deg]
                hex_source.activity = activity_concentration *
hex_source.volume / ui.number_of_threads * gate.g4_units.Bq
                water4sources.append(hex_source)
            # Add the original first pole for i==1
            water4sourcesTubs.append(water_tub)
            source = sim.add_source("GenericSource", f"beta+source_{i}")
            source.particle = "e+"
            source.energy.type = "F18"
            source.energy.mono = 511 * gate.g4_units.keV
            source.position.type = "cylinder"
            source.position.radius = diam * mm
            source.position.dz = 15/2 * mm
            source.position.translation = [x, y, 0 * mm]
            source.volume = np.pi * (source.position.radius ** 2) * 2 *
source.position.dz
            source.energy.mono = 511 * gate.g4_units.keV
            source.direction.type = "iso"
            source.direction.theta = [0 * gate.g4_units.deg, 180 *
gate.g4_units.deg]
            source.direction.phi = [0, 360 * gate.g4_units.deg]
            source.activity = activity_concentration * source.volume /
ui.number_of_threads * gate.g4_units.Bq
            water4sources.append(source)
        # if i == 2:
            # Add green poles at the same radius as the extra red poles (4 *
water_tub.rmax from center)
```

```
        else:
            for j, angle_offset in enumerate([-60, 60]):
                if i== 0 and (j == 0 or j == 1):
                    # Skip the first two extra sources for i==0, they are
already added in the hexagon
                    continue
                extra_tub = sim.add_volume("Tubs",
f"water4source_extra_{i}_{j}")
                extra_tub.mother = phantom.name
                extra_tub.rmax = water_tub_rmax
                extra_tub.rmin = 0 * mm
                extra_tub.dz = water_tub_dz
                angle = angle_offset + angle
                radius = 4 * water_tub_rmax
                x_extra = x + radius * np.cos(np.deg2rad(angle+30))
                y_extra = y + radius * np.sin(np.deg2rad(angle+30))
                extra_tub.translation = [x_extra, y_extra, 0 * mm]
                extra_tub.material = "Water"
                extra_tub.color = colors.red
                water4sourcesTubs.append(extra_tub)
                # Add a source for each extra tub
                extra_source = sim.add_source("GenericSource",
f"beta+source_extra_{i}_{j}")
                extra_source.particle = "e+"
                extra_source.energy.type = "F18"
                extra_source.position.type = "cylinder"
                extra_source.position.radius = water_tub_rmax
                extra_source.position.dz = 15/2 * mm
                extra_source.position.translation = [x_extra, y_extra, 0 * mm]
                extra_source.volume = np.pi * (extra_source.position.radius **
2) * 2 * extra_source.position.dz
                extra_source.energy.mono = 511 * gate.g4_units.keV
                extra_source.direction.type = "iso"
                extra_source.direction.theta = [0 * gate.g4_units.deg, 180 *
gate.g4_units.deg]
                extra_source.direction.phi = [0, 360 * gate.g4_units.deg]
                extra_source.activity = activity_concentration *
extra_source.volume / ui.number_of_threads * gate.g4_units.Bq
                # print(f"extra_source.activity={extra_source.activity}")
                # print(f"extra_source.volume={extra_source.volume}")
                water4sources.append(extra_source)
                # print(f"extra_source.volume={extra_source.volume}")
                water4sources.append(extra_source)
            # For i != 0, add the original first pole
            water4sourcesTubs.append(water_tub)
            if i!=0: source = sim.add_source("GenericSource",
f"beta+source_{i}")
            source.particle = "e+"
```

```
            source.energy.type = "F18"
            source.energy.mono = 511 * gate.g4_units.keV
            source.position.type = "cylinder"
            source.position.radius = diam * mm
            source.position.dz = 15/2 * mm
            source.position.translation = [x, y, 0 * mm]
            source.volume = np.pi * (source.position.radius ** 2) * 2 *
source.position.dz
            source.energy.mono = 511 * gate.g4_units.keV
            source.direction.type = "iso"
            source.direction.theta = [0 * gate.g4_units.deg, 180 *
gate.g4_units.deg]
            source.direction.phi = [0, 360 * gate.g4_units.deg]
            source.activity = activity_concentration * source.volume /
ui.number_of_threads * gate.g4_units.Bq
            # print(f"source.activity={source.activity}")
            # print(f"source.volume={source.volume}")
            water4sources.append(source)
    # Add one magenta tub of radius Tubs_radius_from_diam[2]
    magenta_tub_i2_radius = Tubs_radius_from_diam[2] * mm
    magenta_tub_i2 = sim.add_volume("Tubs", "magenta_tub_i2")
    magenta_tub_i2.mother = phantom.name
    magenta_tub_i2.rmax = magenta_tub_i2_radius
    magenta_tub_i2.rmin = 0 * mm
    magenta_tub_i2.dz = 50/2 * mm
    # build for i=2
    i=2
    angle = i * (360 / len(Tubs_radius_from_diam))
    x = 1*(15+8*magenta_tub_i2_radius*np.sqrt(3))/2 *
np.cos(np.deg2rad(angle)) * mm
    y = 1*(15+8*magenta_tub_i2_radius*np.sqrt(3))/2 *
np.sin(np.deg2rad(angle)) * mm
    magenta_tub_i2.translation = [x, y, 0 * mm]
    magenta_tub_i2.material = "Water"
    magenta_tub_i2.color = colors.magenta

    # Associate a source with the magenta tub
    magenta_source = sim.add_source("GenericSource", "beta+source_magenta")
    magenta_source.particle = "e+"
    magenta_source.energy.type = "F18"
    magenta_source.energy.mono = 511 * gate.g4_units.keV
    magenta_source.position.type = "cylinder"
    magenta_source.position.radius = magenta_tub_i2_radius
    magenta_source.position.dz = 15/2 * mm
    magenta_source.position.translation = magenta_tub_i2.translation
    magenta_source.volume = np.pi * (magenta_source.position.radius ** 2) * 2
* magenta_source.position.dz
    magenta_source.direction.type = "iso"
```

```
    magenta_source.direction.theta = [0 * gate.g4_units.deg, 180 *
gate.g4_units.deg]
    magenta_source.direction.phi = [0, 360 * gate.g4_units.deg]
    magenta_source.activity = activity_concentration * magenta_source.volume /
ui.number_of_threads * gate.g4_units.Bq
    water4sources.append(magenta_source)

    # Add two additional magenta tubs and sources perpendicular to the line
from center to magenta_tub_i2
    # Compute the direction vector from center to magenta_tub_i2
    center_to_magenta = np.array([x, y])
    if np.linalg.norm(center_to_magenta) == 0:
        unit_vec = np.array([1.0, 0.0])
    else:
        unit_vec = center_to_magenta / np.linalg.norm(center_to_magenta)
    # Perpendicular vectors (rotate by 90 deg)
    perp_vec1 = np.array([-unit_vec[1], unit_vec[0]])
    perp_vec2 = np.array([unit_vec[1], -unit_vec[0]])
    sep = 4 * magenta_tub_i2_radius
    for idx, perp_vec in enumerate([perp_vec1, perp_vec2]):
        perp_pos = center_to_magenta + perp_vec * sep
        magenta_tub_perp = sim.add_volume("Tubs",
f"magenta_tub_i2_perp{idx+1}")
        magenta_tub_perp.mother = phantom.name
        magenta_tub_perp.rmax = magenta_tub_i2_radius
        magenta_tub_perp.rmin = 0 * mm
        magenta_tub_perp.dz = 50/2 * mm
        magenta_tub_perp.translation = [perp_pos[0], perp_pos[1], 0 * mm]
        magenta_tub_perp.material = "Water"
        magenta_tub_perp.color = colors.magenta

        # Associate a source with each perpendicular magenta tub
        magenta_source_perp = sim.add_source("GenericSource",
f"beta+source_magenta_perp{idx+1}")
        magenta_source_perp.particle = "e+"
        magenta_source_perp.energy.type = "F18"
        magenta_source_perp.energy.mono = 511 * gate.g4_units.keV
        magenta_source_perp.position.type = "cylinder"
        magenta_source_perp.position.radius = magenta_tub_i2_radius
        magenta_source_perp.position.dz = 15/2 * mm
        magenta_source_perp.position.translation = [perp_pos[0], perp_pos[1],
0 * mm]
        magenta_source_perp.volume = np.pi *
(magenta_source_perp.position.radius ** 2) * 2 *
magenta_source_perp.position.dz
        magenta_source_perp.direction.type = "iso"
        magenta_source_perp.direction.theta = [0 * gate.g4_units.deg, 180 *
gate.g4_units.deg]
```

```
        magenta_source_perp.direction.phi = [0, 360 * gate.g4_units.deg]
        magenta_source_perp.activity = activity_concentration *
magenta_source_perp.volume / ui.number_of_threads * gate.g4_units.Bq
        water4sources.append(magenta_source_perp)
    #print(f"Phantom name: {phantom.name}")
    return phantom
    #, water4sources, water4sourcesTubs
```